\documentclass[a4paper,12pt]{article}
\usepackage{amsmath, amssymb}
\usepackage{authblk}
\usepackage{graphicx}
\usepackage{hyperref}
\usepackage{physics}
\usepackage{cite}

\title{Topology-Enhanced Superconducting Qubit Networks for In-Sensor Quantum Information Processing}

\author[1,7,*]{J. Settino}
\author[2]{G. G. Luciano}
\author[3]{A. Di Bartolomeo}
\author[4,6]{P. Silvestrini}
\author[5]{M. Lisitskiy}
\author[6]{B. Ruggiero}
\author[3,8]{F. Romeo}

\affil[1]{Dipartimento di Fisica, Università della Calabria, Rende (CS), I-87036, Italy}

\affil[2]{Departamento de Qu\'{\i}mica, F\'{\i}sica y Ciencias Ambientales y del Suelo, Escuela Polit\'ecnica Superior -- Lleida, Universidad de Lleida, Av. Jaume II, 69, 25001 Lleida, Spain}

\affil[3]{Dipartimento di Fisica ”E. R. Caianiello”,
Università degli Studi di Salerno,
Via Giovanni Paolo II, I-84084 Fisciano (SA), Italy}

\affil[4]{Dipartimento di Matematica e Fisica, Università della Campania “Luigi Vanvitelli”, Viale Abramo
Lincoln 5, I-81100 Caserta, Italy}

\affil[5]{CNR-SPIN, Institute for Superconductivity, Innovative Materials and Devices, I-80078 Pozzuoli (NA), Italy}

\affil[6]{CNR-ISASI, Institute of Applied Sciences and Intelligent Systems, Via Campi Flegrei 34, Building 70, I-80078 Pozzuoli (NA), Italy}

\affil[7]{INFN, Gruppo Collegato di Cosenza, I-87036 Arcavacata di Rende (CS), Italy}

\affil[8]{INFN, Sezione di Napoli, Gruppo Collegato di Salerno, Via Giovanni Paolo II, I-84084 Fisciano (SA), Italy}

\affil[*]{Corresponding author: jacopo.settino@unical.it}
\date{\today}
\begin{document}
\maketitle

\begin{abstract}
We investigate the influence of topology on the magnetic response of inductively coupled superconducting flux‑qubit networks. Using exact diagonalization methods and linear response theory, we compare the magnetic response of linear and cross‑shaped array geometries, used as paradigmatic examples. We find that the peculiar coupling matrix in cross‑shaped arrays yields a significant enhancement of the magnetic flux response compared to linear arrays, this network-topology effect arising from cooperative coupling among the central and the peripheral qubits. These results establish quantitative design criteria for function-oriented superconducting quantum circuits, with direct implications for advancing performance in both quantum sensing and quantum information processing applications. Concerning the latter, by exploiting the non-linear and high-dimensional dynamics of such arrays, we demonstrate their suitability for quantum reservoir computing technology. This dual functionality suggests a novel platform in which the same device serves both as a quantum-limited electromagnetic sensor and as a reservoir capable of signal processing, enabling integrated quantum sensing and processing architectures.

\end{abstract}

\section{Introduction}
Quantum systems exhibiting coherence on nanoscopic scales have become pivotal in contemporary nanoelectronics research\cite{Heinrich2021}, driving significant technological advancements\cite{Salhov2024}. Among these, superconducting electronic systems\cite{Barone1982,DeLuca2020} represent an important class of platforms capable of reliably maintaining and manipulating quantum coherence\cite{Silvestrini1996,Ruggiero2003, Mooij1999,Makhlin2001,Blais2004,Koch2007,Clarke2008,Devoret2013,Wendin2017,Gu2017,Kjaergaard2020}. Flux qubits\cite{Orlando1999,Chiorescu2003,Yan2016,Macha2014}, in particular, have emerged as one of the leading technologies in the pursuit of quantum computation, leveraging the coherent superposition of macroscopic circulating currents\cite{VanderWal2000} as fundamental elements for quantum information processing.

A central challenge is the realization of large-scale qubit networks\cite{Boixo2014,King2018} without incurring prohibitive levels of decoherence\cite{Hanzo2025}, which would severely degrade the functionality of quantum devices. Addressing conditions that ensure resilience to decoherence is crucial, not only to enhance computational performance but also to exploit these networks as highly sensitive quantum sensors\cite{Degen2017,Poggiali2018,Ruggiero1999,Kantsepolsky2023}. In addition to material research for the optimization of the coherence parameter of quantum devices \cite{misha1,misha2}, open questions also remain regarding optimal strategies for controlling interactions among superconducting qubits, with major challenges arising from the need to scale quantum networks without compromising coherence.

Concerning quantum sensing applications, the dual role of flux-qubit networks as both a quantum-limited electromagnetic sensor\cite{Ilichev2007} and a quantum reservoir \cite{Fujii2017,Mujal2021,Ghosh2021,Martínez2021,Innocenti2023,Bravo2022} capable, for instance, of performing inverse problem solving becomes particularly compelling. Indeed, by monitoring suitable observables of the network in response to an external signal, one may not only detect the signal with high sensitivity but also reconstruct it through in-sensor quantum information processing. The present study provides a necessary step toward this vision, by analyzing the static and dynamical response properties of qubit networks in different topological configurations. These foundational results are essential for assessing the expressive capabilities of the system and for enabling future architectures that integrate quantum sensing with quantum reservoir computing.

To follow our program, we analyze the extent to which the topology of inductive couplings, as defined in graph theory, affects the emergent physical properties of flux-qubit networks. We consider, in particular, two representative architectures: a linear array (LA)\cite{Fistul2022} and a cross-shaped array (CA). Theoretical analysis reveals that the structure of the coupling network critically influences the excitation spectrum, in close analogy with concepts from graph theory and complex networks. This insight is consistent with previous theoretical\cite{Burioni2000,Lorenzo2014,Romeo2021,Romeo2022,Romeo2023} and experimental\cite{Bizzi2023,Lucci2022,Silvestrini2007,Lucci2020} findings showing that the dynamics of coherent quantum systems defined on graphs is strongly shaped by the presence of highly connected nodes, which act as hubs mediating collective behavior.

The paper is thus organized as follows. After the introduction of the fundamental physics of the isolated flux qubit (Sec.~\ref{sec:2}) and of flux qubit networks (Sec.~\ref{sec:3}), we introduce the \textit{coherence-assisted topological enhancement paradigm} in Sec.~\ref{sec:4}, and we investigate in Sec.~\ref{sec:5} the static properties of arrays in two distinct geometries (i.e., LA and CA), emphasizing the critical role played by topology in shaping the pattern of circulating currents. The magnetic flux generated by arrays with different topological configurations and its potential detection are issues addressed in Sec.~\ref{sec:6}. In Sec.~\ref{sec:7}, we turn to the analysis of qubit arrays coupled to a transmission line. By combining exact diagonalization techniques with linear response theory, we provide a detailed characterization of the dynamical response of the arrays to external perturbations. The analysis of the arrays response unambiguously demonstrates that both spectral and dynamical properties are strongly governed by the underlying topological structure, offering key insights for the design and optimization of topologically-engineered quantum devices. Finally, in Sec.~\ref{sec:8}, we test the expressive capabilities of a quantum reservoir computer based on a superconducting qubit network in a prototypical prediction task and present our conclusions in Sec.~\ref{sec:9}.

\section{Physics of an Isolated Flux Qubit}\label{sec:2}
\begin{figure*}
	\centering
	\includegraphics[width=1\textwidth]{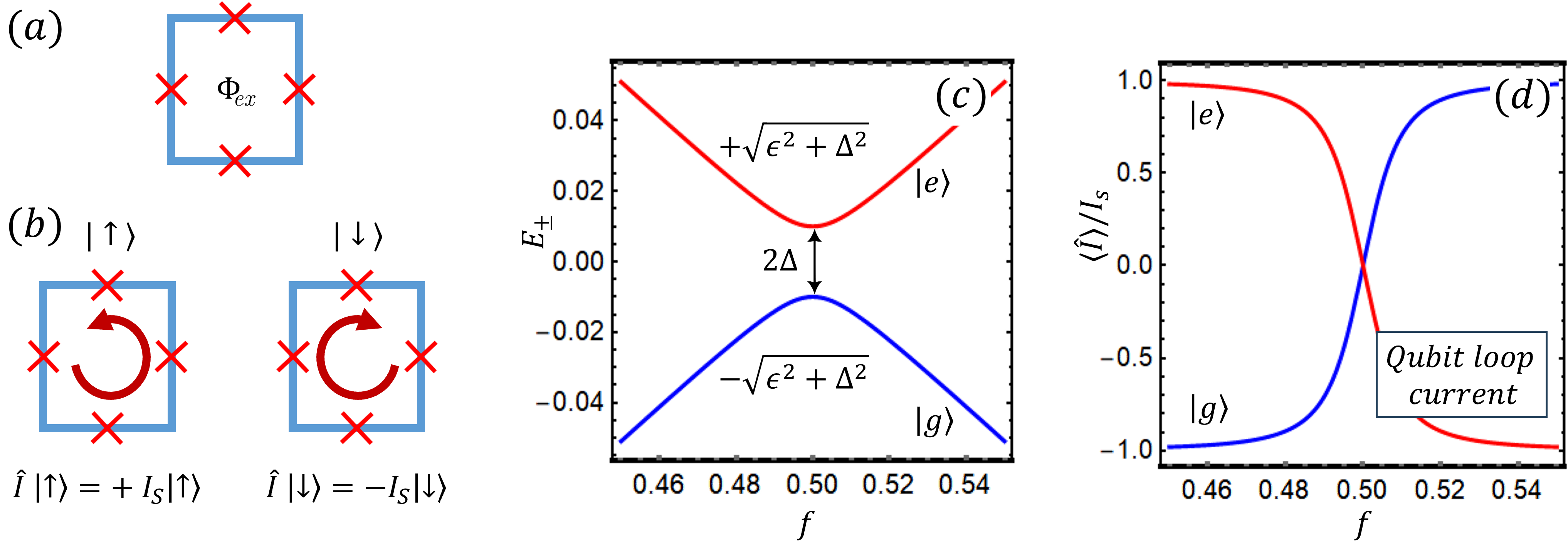}
	\caption{%
			(a) Schematic of a flux qubit realized by means of the four Josephson junctions technology \cite{Qiu2016}. The red crosses denote the weak links between superconducting regions (blue lines). 
			(b) Circulating currents in the qubit loop: counterclockwise and clockwise currents correspond to the current states \( |\uparrow\rangle \) and \( |\downarrow\rangle \), respectively. 
			(c) Energy spectrum of the effective Hamiltonian \( H_{\text{eff}} \) as a function of the normalized magnetic flux \( f \). 
			(d) Expectation value of the current operator on the energy eigenstates, normalized to the maximum loop current $I_s$ and plotted as a function of \( f \).%
			}
	\label{F1}
\end{figure*}

As a preliminary step towards the analysis of qubit arrays, we review the fundamental properties of an isolated flux qubit (Fig. \ref{F1}(a)), introducing the relevant notation and key flux-dependent quantities that will be used throughout.
The quantum dynamics of a superconducting flux qubit can be described by an effective Hamiltonian of the form\cite{Mooij1999,Orlando1999,VanderWal2000,Makhlin2001,Clarke2008}:
\begin{equation}
\hat{H}_{\text{eff}} = -[\epsilon(f) \sigma_z + \Delta \sigma_x],
\end{equation}
where $\epsilon(f)$ depends on the normalized external flux $f = \Phi_{ex}/\Phi_0$ according to
\begin{equation}
\epsilon(f) = I_S \Phi_0 \left( f - \frac{1}{2} \right),
\end{equation}
and $\Delta$ is the tunneling energy, which is related to the Josephson and the charging energy of the Josephson junctions constituting the flux qubit. The effective two-level model is written in terms of Pauli matrices $\sigma_x$ and $\sigma_z$ and contains information about the maximum loop supercurrent $I_S$ and the external magnetic flux $\Phi_{ex}$, compared to the magnetic flux quantum $\Phi_0=h/(2e)$.
The current operator associated with the qubit loop is defined as\cite{Wu2013}:
\begin{equation}
\hat{I} = -\frac{\partial \hat{H}_{\text{eff}}}{\partial \Phi_{ex}} = I_S \sigma_z.
\end{equation}
Introducing the eigenstates $\{ |\uparrow\rangle, |\downarrow\rangle \}$ of the Pauli matrix $\sigma_z$, defined by the relations:
\begin{equation}
\sigma_z |\uparrow\rangle = + |\uparrow\rangle, \quad \sigma_z |\downarrow\rangle = - |\downarrow\rangle,
\end{equation}
we observe that:
\begin{equation}
\hat{I} |\uparrow\rangle = +I_S |\uparrow\rangle, \quad \hat{I} |\downarrow\rangle = -I_S |\downarrow\rangle.
\end{equation}
From a physics point of view, the state $|\uparrow\rangle$ represents a counterclockwise circulating current, whereas the state $|\downarrow\rangle$ represents a clockwise current (Fig. \ref{F1}(b)).
The Hamiltonian can be diagonalized, yielding eigenvalues (Fig. \ref{F1}(c)):
\begin{equation}
E_{\pm} = \pm \sqrt{\epsilon^2 + \Delta^2},
\end{equation}
and corresponding eigenstates:
\begin{equation}
|g\rangle = \sqrt{\frac{1 + \cos(\theta)}{2}} |\uparrow\rangle + \sqrt{\frac{1 - \cos(\theta)}{2}} |\downarrow\rangle,
\end{equation}
\begin{equation}
|e\rangle = \sqrt{\frac{1 - \cos(\theta)}{2}} |\uparrow\rangle - \sqrt{\frac{1 + \cos(\theta)}{2}} |\downarrow\rangle,
\end{equation}
where:
\begin{equation}
\cos(\theta) = \frac{\epsilon}{\sqrt{\epsilon^2 + \Delta^2}}.
\end{equation}
The expectation values of the current in the energy eigenstates are:
\begin{equation}
\langle \hat{I} \rangle_g = I_S \cos(\theta), \quad \langle \hat{I} \rangle_e = -I_S \cos(\theta).
\end{equation}
Thus, the energy eigenstates are a coherent superposition of states carrying opposite circulating currents, whose weight depends on the angle $\theta$. The latter, in turn, is controlled by the physical parameters of the qubit and, in particular, by the external flux.\\
Interestingly, $\cos(\theta)$ is a vanishing quantity when the external flux is equal to half the flux quantum  (i.e., $f=1/2$). Under this condition, the energy eigenstates are symmetric and antisymmetric combinations of the form 
\begin{eqnarray}
(|\uparrow \rangle \pm |\downarrow \rangle)/\sqrt{2},
\end{eqnarray}
while the loop current expectation values take vanishing values (i.e. $\langle \hat{I} \rangle_{g/e}=0$).\\ 
The externally applied magnetic flux \( \Phi_{\mathrm{ex}} \) acts as a tunable control parameter for the flux qubit, governing the direction of the persistent supercurrent circulating in the superconducting loop. This behavior becomes evident when analyzing the expectation value of the ground-state current as a function of \( \Phi_{\mathrm{ex}} \) (Fig. \ref{F1}(d)). In response to an applied flux, the system establishes a circulating current that adjusts the total magnetic flux threading the loop toward quantized values, either \( 0 \) or \( \Phi_0 \), corresponding to the minima of an effective double-well potential.

Specifically, when \( \Phi_{\mathrm{ex}} > \Phi_0/2 \), the ground state is characterized by a counterclockwise supercurrent, which generates a positive magnetic flux. In contrast, for \( \Phi_{\mathrm{ex}} < \Phi_0/2 \), the energy minimum corresponds to a clockwise supercurrent, which yields a negative flux contribution. These currents originate self-induced magnetic fluxes that enforce the quantization condition for the total flux, which stabilizes around \( \Phi = n_f \Phi_0 \), with \( n_f \in \{0,1\} \).

\begin{figure}
	\centering
	\includegraphics[width=1.0\textwidth]{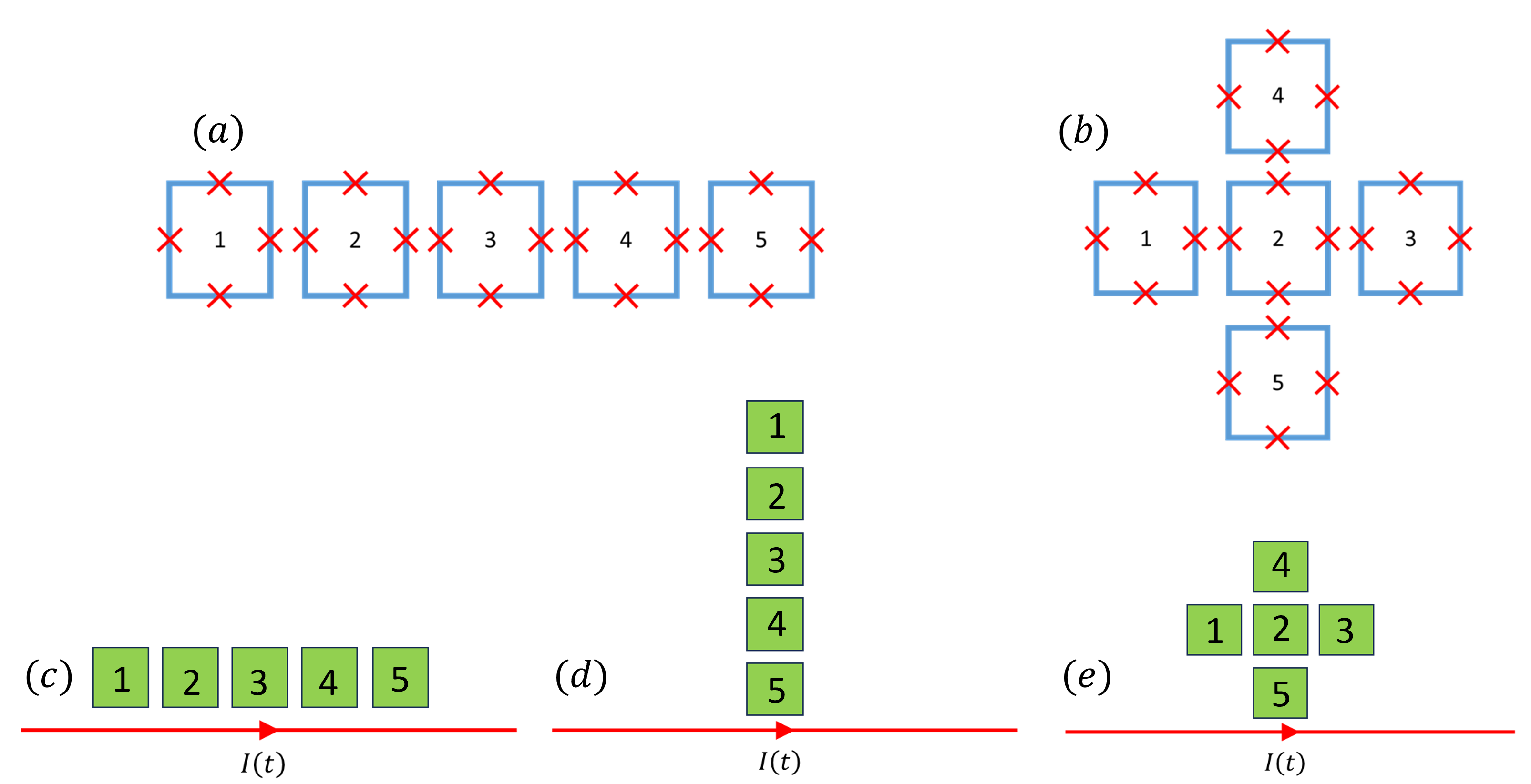}
	\caption{Schematic of a five-qubit array with inductive coupling: (a) linear configuration and (b) cross-shaped configuration. (c)-(e) Different coupling conditions with a transmission line, biased by the current $I(t)$.}
	\label{F2}
\end{figure}

\section{Superconducting Qubit Networks}
\label{sec:3}
The Hamiltonian of a collection of superconducting flux qubits that interact via inductive coupling (see for instance Fig. \ref{F2}, panel (a) and (b)) is given by\cite{Fistul2022}:
\begin{equation}\label{eq:qubitNetwork}
H = \sum_i H_Q^{(i)} + \frac{1}{2} \sum_{\substack{i,j \\ i \neq j}} U_{ij},
\end{equation}
where the local qubit term is:
\begin{equation}
H_Q^{(i)} = -\left[ \epsilon_i(f) \sigma_z^{(i)} + \Delta_i \sigma_x^{(i)} \right],
\end{equation}
while the interaction between qubits is:
\begin{equation}
U_{ij} = M_{ij} I_S^{(i)} I_S^{(j)} \sigma_z^{(i)} \sigma_z^{(j)},
\end{equation}
with $M_{ij}$ the mutual inductance coefficient between qubits $i$ and $j$, while $I_S^{(k)}$ represents the maximum loop current which can be sustained by the $k$-th qubit. For a single-junction flux qubit, the maximum loop current coincides with the junction critical current.

If the qubit network is coupled to a transmission line (Fig. \ref{F2}, panels (c)-(e)) carrying a time-dependent current $I(t)$, an additional mutual-inductance contribution enters the Hamiltonian\cite{Fistul2022}:
\begin{equation}
H_{I} = \sum_i \mathcal{M}_{i} I(t) I_S^{(i)} \sigma_z^{(i)}.
\end{equation}
By properly modulating $I(t)$, this interaction allows both the control of the quantum state of the network and the induction of coherent transitions between its energy levels.

\section{Coherence-Assisted Topological Enhancement Paradigm
}\label{sec:4}
The present analysis is motivated by a line of research that has progressively unveiled the fundamental role of the interplay of topology and coherence in cold bosonic atoms and in superconducting systems defined on graph-like structures.\\
Indeed, from a theoretical standpoint, Refs. \cite{Burioni2000,Lorenzo2014,Romeo2021,Romeo2022,Romeo2023} have shown that inhomogeneous connectivity can induce nontrivial collective phenomena such as Bose–Einstein condensation on star and comb graphs or the emergence of topological focusing in discrete Ginzburg–Landau models. In these systems, nodes with higher degree act as attractive centers for the condensate or for the superconducting order parameter.\\
These studies demonstrated that the network topology itself can promote spatial localization and amplification phenomena, effectively turning connectivity into a control parameter of the macroscopic quantum state.\\
Theoretical predictions have subsequently found experimental confirmation in Josephson-junction networks \cite{Bizzi2023,Lucci2022,Silvestrini2007,Lucci2020}, where coherent and topological effects have been directly observed through critical current enhancement and long-range phase correlations. These experiments have established a clear connection between graph topology and superconducting coherence in realistic architectures of connected superconducting islands.\\
Analogous effects also arise in the discrete Schr\"odinger equation on graphs, whereas they are absent in purely classical diffusive processes, where the stationary state is homogeneous and the topology affects only the relaxation dynamics, not the final equilibrium configuration. This distinction emphasizes that topological effects are relevant exclusively to wave-like, coherent systems, in which nontrivial interference patterns sustain organized spatial structures.\\
Building on these theoretical and experimental advances, our goal is to extend the concepts of coherence-assisted topological enhancement to superconducting qubit networks. In this context, we aim to investigate how similar mechanisms can be harnessed within architectures of coupled flux qubits, with potential technological implications for quantum sensing and in-sensor quantum information processing.\\
To explore the interplay between network topology and coherence, we focus below on two paradigmatic five-qubit architectures: a linear array and a cross-shaped array. From a graph-theoretical perspective, a linear array is described by a tridiagonal adjacency matrix in which the node connectivity varies between one and two, depending on whether the node is terminal or internal. In contrast, a cross-shaped array includes a central node with connectivity four and four peripheral nodes with connectivity one. These distinct topological features are expected to yield markedly different collective behaviors. The purpose of the following analysis is to verify that topological enhancement, previously demonstrated and experimentally confirmed in other superconducting platforms, also manifests in flux-qubit networks.

\section{Spectral Properties and Loop-Current Distribution of Linear and Cross-Shaped Arrays}
\label{sec:5}
In line with the program outlined above, we proceed by examining the energy spectrum and loop-current distribution of two distinct five-qubit superconducting networks, namely a linear array and a cross-shaped arrangement, whose qubits are coupled exclusively via mutual inductance. Because a counterclockwise loop current in one qubit generates a negative magnetic flux in its neighbors, the mutual inductance coefficients are taken as negative (\(M_{ij}<0\)). Couplings beyond nearest neighbors are neglected, since the magnetic interaction between distant loops only provides small corrections. Throughout the simulations, the reduced magnetic flux is fixed at \(f=0.52\), i.e., slightly above the optimal working point \(f=1/2\) of the qubit.

\begin{figure}
	\centering
	\includegraphics[width=1.0\textwidth]{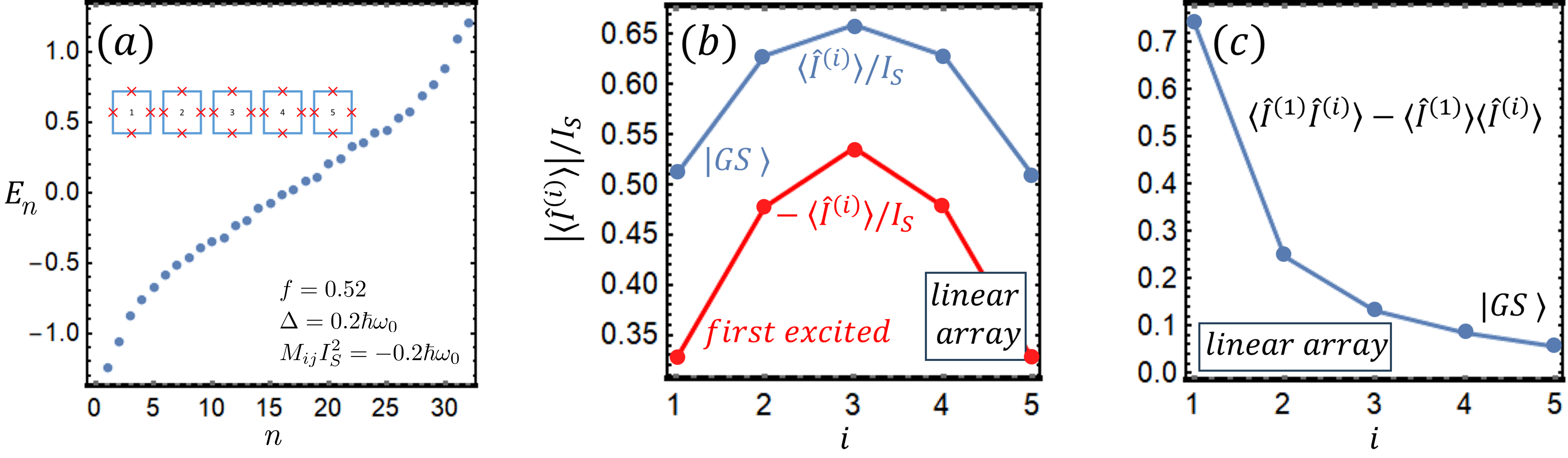}
	\caption{%MODIFICATO
		(a) Energy eigenvalues $E_n$ of a linear array of five identical qubits, labeled by the index $n$, and sorted in ascending order. Energy eigenvalues are expressed in units of $\hbar \omega_0=I_s \Phi_0$. (b) Loop currents of the linear array normalized to the maximum current $I_s$ as a function of the qubit index \( i \) for the ground state and the first excited state. (c) Current-current correlations, in units of $I_S^2$, between the first qubit and the \( i \)-th qubit.%
		}
	\label{F3}
\end{figure}

Figure~\ref{F3}(a) shows the energy spectrum of the linear array obtained by exact diagonalization of the Hamiltonian:
\begin{eqnarray}
H_{LA}=-\sum_{i=1}^{5}[\epsilon(f) \sigma_z^{(i)}+\Delta \ \sigma_{x}^{(i)}]+MI_{S}^{2}\sum_{i=1}^{4} \sigma_z^{(i)} \sigma_z^{(i+1)},
\end{eqnarray}
with model parameters given in the figure inset. The spectrum exhibits a non-degenerate energy-level structure (consistent with the one-dimensional irreducible representations of the $C_2$ symmetry group), with the lowest states well separated from the excited states, reflecting the collective nature of the excitations. In the ground state, the loop currents \( \langle \hat{I}^{(i)} \rangle \) are mirror symmetric with respect to the center of the chain [Fig.~\ref{F3}(b)], with the maximal current amplitude observed in the central qubit. This distribution is a direct consequence of the array topology and the nature of the inductive couplings. The current profile associated with the first excited state displays a similar pattern, even though the loop currents exhibit clockwise circulation and reduced magnitude.

The spatial structure of the current-current correlations, shown in Fig.~\ref{F3}(c), highlights the decay of the quantum correlations with distance. In particular, the current-current correlation function \( \mathcal{C}(1,i)=\langle \hat{I}^{(1)} \hat{I}^{(i)} \rangle - \langle \hat{I}^{(1)} \rangle \langle \hat{I}^{(i)} \rangle \) decreases monotonically as the index \( i \) increases. This behavior reveals how the inductive coupling mediates genuine long-range quantum correlations across the array.

\begin{figure}
	\includegraphics[width=1.0\textwidth]{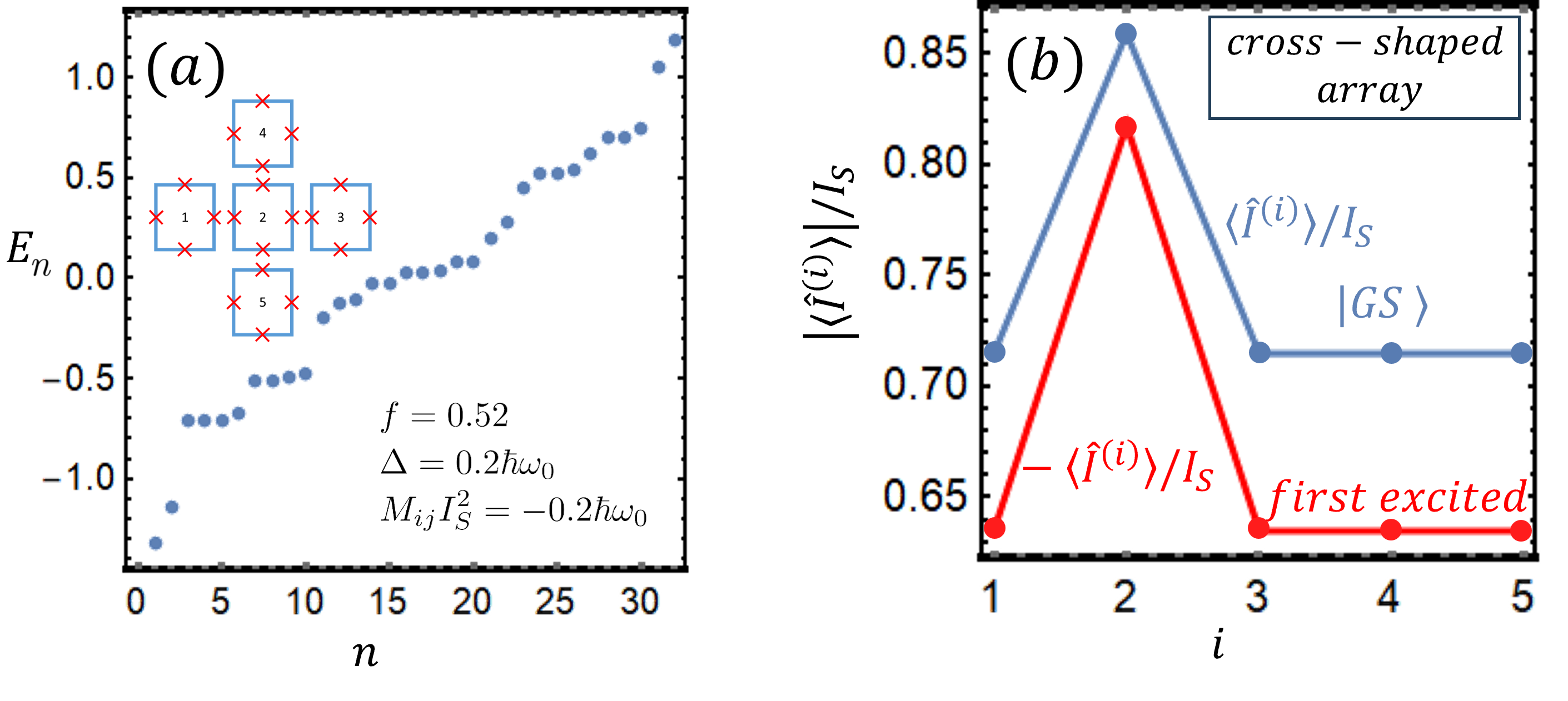}
	\centering
	\caption{%MODIFICATO
		(a) Energy eigenvalues $E_n$ of a cross-shaped array of five identical qubits, labeled by the index $n$, and sorted in ascending order. Energy eigenvalues are expressed in units of $\hbar \omega_0=I_s \Phi_0$. (b) Loop currents of the cross-shaped array normalized to the maximum current $I_s$ as a function of the qubit index \( i \) for the ground state and the first excited state.%
	}
	\label{F4}
\end{figure}

The energy spectrum of the cross-shaped array, derived from the Hamiltonian:
\begin{eqnarray}
H_{CA}=-\sum_{i=1}^{5}[\epsilon(f) \sigma_z^{(i)}+\Delta \ \sigma_{x}^{(i)}]+MI_{S}^{2}\ 
\sigma_z^{(2)} \sum_{i\neq2} \sigma_z^{(i)} ,
\end{eqnarray}
is reported in Fig.~\ref{F4}(a). In this configuration, qubit 2 occupies the central position and is inductively coupled to the four peripheral qubits. Due to the \(C_{4v}\) point-group symmetry (90$^\circ$ rotations and mirror lines) of the array, the low-energy sector of the energy spectrum organizes into two groups of nearly generate levels (corresponding to states presenting both accidental quasi-degeneracy and symmetry-protected degeneracy), which play a relevant role in defining an evident energy gap between the first excited state and the second excited state. This symmetry-induced phenomenon, not observed in the linear array case, appears to be relevant in shaping the response of a quantum device with function-oriented design.
Moreover, the enhanced connectivity of this geometry significantly alters the current distribution. Indeed, as shown in Fig.~\ref{F4}(b), the ground-state current profile is now highly localized at the central qubit, which carries the largest loop current. This behavior is also observed in the first excited state, where the current enhancement factor is even more pronounced compared to the one observed in the ground state.

The marked difference between the low-energy spectra of the linear and cross-shaped arrays highlights the crucial role that network topology plays in determining the emergent properties of the system.  
In particular, the cross-shaped array exhibits a marked enhancement of the central loop current, consistent with the topological focusing mechanism discussed in Sec.~\ref{sec:4} and reminiscent of the localization effects previously discussed within the Ginzburg–Landau framework in Ref. \cite{Romeo2021}.
This amplification can be viewed as a cooperative effect induced by network connectivity and can also be interpreted within the framework of spectral graph theory.
In this context, the nearest-neighbor inductive couplings define the adjacency matrix of an undirected graph, in which each qubit represents a node.
Highly connected nodes act as hubs that govern the collective response of the system, serving as focal points for magnetic flux concentration and thus shaping the overall properties of the quantum network.

The spectral properties of the cross-shaped array suggest another promising perspective consisting of encoding a logical qubit in the many-body manifold spanned by the collective ground state \(\lvert G\rangle\) and the first excited state \(\lvert E\rangle\) of the entire cross-shaped array. In view of the presented spectral properties, this choice may passively suppress uncorrelated decoherence through symmetry-protected subradiance.  The expected advantage over a single qubit complemented by active error correction must be validated experimentally by charting the regimes of temperature, fabrication tolerance, and noise correlation where this protection proves effective.

Thus, the ability to control the graph structure offers a promising route for tailoring quantum responses in superconducting qubit networks.

\section{Static Magnetic Response of a Superconducting Qubit Network}
\label{sec:6}
The current pattern in qubit networks is responsible for a magnetic flux, which can be measured via a sensing loop. As the expectation value of \( \hat{\sigma}_z^{(i)} \) is proportional to the persistent current circulating in the \( i \)-th qubit, the flux \( \Phi \) through a nearby sensing loop is given, up to a constant prefactor, by

\begin{equation}
    \Phi \propto \sum_i \langle \hat{\sigma}_z^{(i)} \rangle_{\mathrm{GS}}.
\end{equation}

The ground-state expectation value is justified by the experimental conditions, as the system is typically operated at millikelvin temperatures and in a low-noise electromagnetic environment, ensuring negligible thermal excitations.

\begin{figure}[t]
    \centering
    \includegraphics[width=0.9\linewidth]{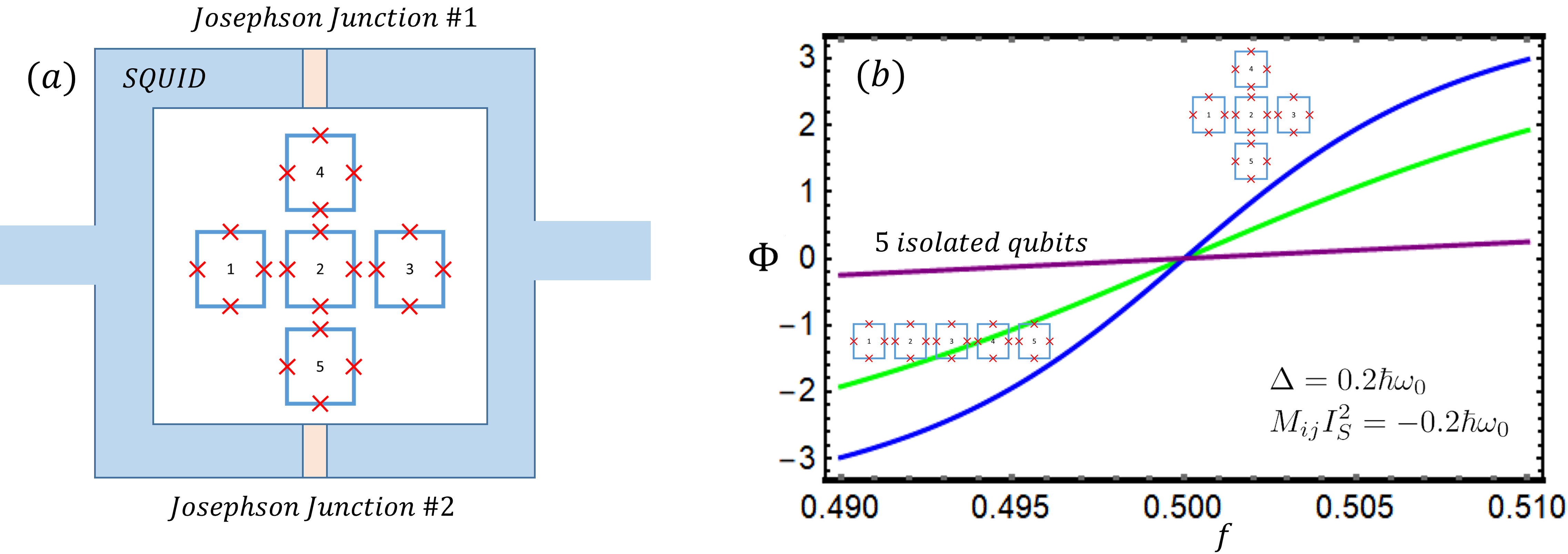}
    \caption{(a) Qubit array embedded in a dc SQUID, used as a flux sensor. (b) Magnetic flux \( \Phi\) generated by the qubit network as a function of the normalized external flux \( f \) threading each qubit. A comparison is performed among three different qubit configurations: isolated, linear- and cross-shaped array.}
    \label{F5}
\end{figure}

Figure~\ref{F5}(a) shows a schematic of a possible detection setup, where the qubit array is embedded within a d.c. SQUID, used as a flux sensor enabling non-invasive measurement of the flux generated by the qubit network. The magnetic flux generated by the qubit network in response to the external flux \( f \) is shown in Fig.~\ref{F5}(b) for three different configurations: five non-interacting qubits, a linear array and a cross-shaped array. In all cases, the physical parameters are kept fixed (see the inset of Fig. \ref{F5}(b)).

The simulation results clearly demonstrate that for identical local parameters and applied flux values the cross-shaped configuration produces the largest flux response. This enhancement reflects the coherent buildup of loop currents in the presence of higher connectivity. In contrast, the flux signal from isolated qubits grows linearly with \( f \), while the linear array exhibits intermediate behavior.

These findings confirm that the topology of the qubit network plays a central role in shaping the measurable electromagnetic response. From a practical perspective, the stronger signal associated with the cross-shaped geometry suggests an evident strategy for optimizing qubit readout architectures. Topological engineering may indeed serve as a powerful resource in enhancing the detectability of quantum states in superconducting circuits.

\section{Response of a Qubit Network to Dynamical Coupling with a Transmission Line}
\label{sec:7}
To investigate the influence of an external electromagnetic environment, we consider the coupling between a transmission line and the qubit array. The resulting response of the system is evaluated within linear response theory, based on a perturbative treatment of the interaction Hamiltonian. This approximation is justified when either the system is subject to weak signals or it operates in the weak-coupling regime. We develop the theory at zero temperature, consistently with the millikelvin regime in which these systems are usually experimentally probed. Proceeding in the standard way\cite{Fetter2012}, we can write the Schr\"odinger equation governing the quantum evolution as follows:
\begin{eqnarray}
i \hbar \partial_t |\psi_S(t)\rangle = [H_0 + H_{I}(t)]|\psi_S(t)\rangle,
\end{eqnarray}
where $H_0$, identical to $H$ in Eq.~(\ref{eq:qubitNetwork}), represents the Hamiltonian of the generic flux-qubit network. %MODIFICATO
The coupling term with the transmission line can be expressed as $H_{I}(t)=f(t) \ \theta(t-t_0) \ B_{S}$, where $f(t)$ is a time-dependent function proportional to the current $I(t)$ flowing in the transmission line, $B_{S}$ is an appropriate operator in Schr\"odinger representation, while the Heaviside step function $\theta(t-t_0)$ ensures that the coupling is activated only after the onset time $t_0$. The expectation value of a generic operator $A_{S}$ in Schr\"odinger representation, purified from contributions not related to the external perturbation, is given by the convolution integral:
\begin{eqnarray}
\delta \langle A(t) \rangle= \int_{t_0}^{\infty} \chi(t-t')f(t')dt',
\end{eqnarray}
expressed in terms of the response function kernel:
\begin{eqnarray}
\chi(t-t')=-\frac{i}{\hbar}\theta(t-t')\langle 0|[A_{H}(t),B_{H}(t')]|0\rangle.
\end{eqnarray}
The response function $\chi(t-t')$ is written in terms of operators in the Heisenberg representation, i.e.
\begin{eqnarray}
A_{H}(t)=\exp{(iH_0 t/\hbar)}A_S \exp{(-iH_0 t/\hbar)},
\end{eqnarray}
and
\begin{eqnarray}
B_{H}(t)=\exp{(iH_0 t/\hbar)}B_S \exp{(-iH_0 t/\hbar)},
\end{eqnarray}
and contains information about the ground state $|0 \rangle$ of the array in the absence of coupling to the transmission line. Under the assumption that the perturbation is turned on in the remote past ($t_0 \rightarrow -\infty$), the response function can be written in the spectral representation as follows\cite{Fetter2012}:
\begin{eqnarray}
\label{eq:chi}
\chi(\omega)=\sum_n \Bigl\{\frac{\langle0|A_{S}|n \rangle \langle n|B_{S}|0 \rangle}{\hbar \omega-(E_n-E_0)+i \eta}-\frac{\langle0|B_{S}|n \rangle \langle n|A_{S}|0 \rangle}{\hbar \omega+(E_n-E_0)+i \eta}\Bigr\},
\end{eqnarray}
where $\eta$ is an infinitesimal quantity and $E_n$ is the energy eigenvalue associated with the eigenstate $|n\rangle$ of the unperturbed array defined by $H_0 |n\rangle=E_n |n\rangle$. Weak dissipative effects are incorporated by assigning a small but finite value to the parameter $\eta$, which emulates the coupling to the measurement environment. In studying the system's response, it is important to determine the magnetic flux generated by the array in response to the external perturbation. For this purpose, for instance, we set $A_S=B_S=\sum_i \sigma_z^{(i)}$ to model the response function of a linear array coupled to a transmission line as shown in Fig.\ref{F2}(c). Once the response function is known, the time-dependent component of the flux generated by the array in response to the perturbation induced by the transmission line can be calculated as
\begin{eqnarray}
\delta \langle A(t) \rangle = \mathcal{A}|\chi(\omega)| \sin (\omega t-\phi(\omega)),
\end{eqnarray}
where we have chosen $f(t)=\mathcal{A} \sin (\omega t)$, with $\mathcal{A}$ being a constant that absorbs the relevant physical quantities for the coupling between the transmission line and the array. The quantity $\chi(\omega)$ can be experimentally estimated by analyzing the transmission and reflection properties of the transmission line as a function of frequency $\omega$ (see, for instance, Ref. \cite{Fistul2022}), making this response function significant for the theoretical and experimental description of the flux response properties of the array. The function $\chi(\omega)$ exhibits poles at frequencies $(E_n-E_0)/\hbar$, corresponding to transitions between the ground state of the array and any of its excited states due to external perturbation. The visibility of resonances in $\chi(\omega)$ at frequencies $(E_n-E_0)/\hbar$ requires non-vanishing values of the quantity $\langle 0|A_S |n \rangle \langle n|B_S |0 \rangle$. Thus, within the framework of linear response theory, the response of the system is fully characterized by the spectral properties of the isolated array. These properties strongly depend on the array topology, so that tailored responses to external signals can be obtained by modifying the inductive coupling between the qubits.

\subsection{Dynamical Response of a Linear Array}
\label{sec: LA dynamic}
For a linear array of identical flux qubits, the homogeneous coupling to a transmission line is captured by the interaction Hamiltonian (see Fig. \ref{F2}(c)):
\begin{equation}
H_I = \mathcal{M} I_S I(t) \sum_i \sigma_z^{(i)} \equiv f(t) \sum_i \sigma_z^{(i)}.
\end{equation}
Consequently, the flux response of the qubit network to the external signal can be characterized by analyzing the susceptibility \(\chi(\omega)\) in Eq.~(\ref{eq:chi}), with $A_S = B_S = \sum_i \sigma_z^{(i)}$. As a direct consequence of the latter expression, the occurrence of quantum transitions between the ground state and the $n$th excited state is governed by the quantity
$|\langle 0|\sum_i \sigma_z^{(i)}|n \rangle |^2$, which determines the strength of resonant peaks in $\chi(\omega)$ and thus shapes the overall dynamical response of the system. The Hamiltonian of the isolated qubit network can be diagonalized to obtain its spectral properties, which are subsequently used to evaluate the response function $\chi(\omega)$. Once $\chi(\omega)$ has been obtained, the magnetic response of the system can be characterized as a function of the external parameters and, in particular, under different values of the applied static magnetic flux $f=\Phi_{ex}/\Phi_0$.\\

\begin{figure}
	\centering
	\includegraphics[width=1.0\textwidth]{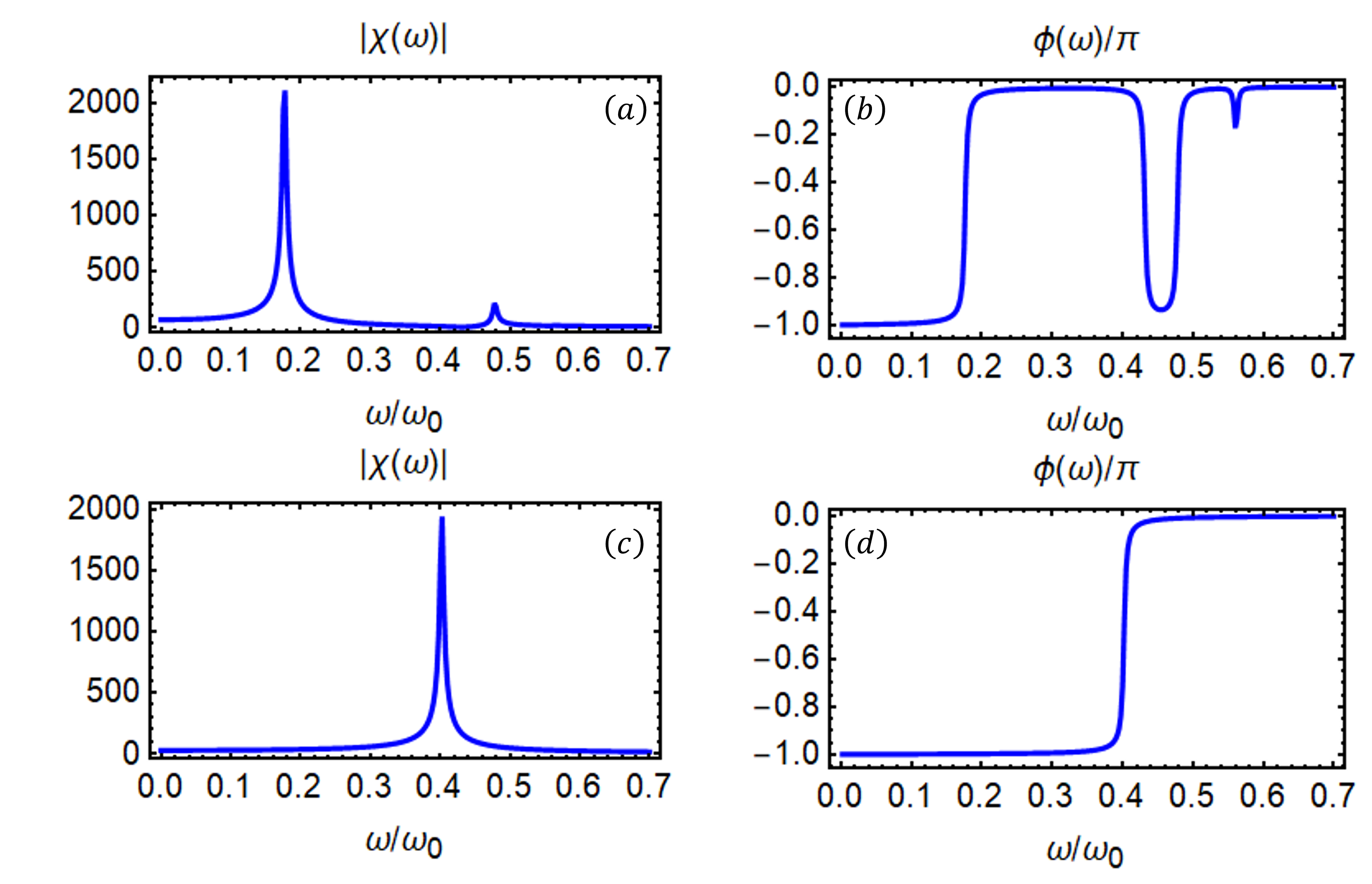}
	\caption{\textbf{(a)} Response function amplitude $|\chi(\omega)|$ of a linear array of five inductively coupled qubits, plotted versus the normalized drive frequency $\omega/\omega_{0}$ (model parameters: $\Phi_{ex}/\Phi_0=0.52$, $\Delta=-MI_S^2=0.2 \ \hbar \omega_0$, $\eta=2.5 \cdot 10^{-3} \ \hbar \omega_0$, with $\omega_0=I_s \Phi_0/\hbar$). 
    \textbf{(b)} Corresponding response function phase $\phi(\omega)/\pi$. 
    \textbf{(c,d)} Same quantities for the uncoupled case (mutual inductance coupling energy $-MI_s^2$ set to $10^{-6} \ \hbar \omega_0$), i.e. five isolated qubits. 
    Comparison of (a,b) with (c,d) highlights the emergence of multiple resonances and phase discontinuities induced by the qubit–qubit coupling, in contrast to the single resonance and simple $\pi$–phase roll exhibited by uncoupled qubits.}
	\label{F6}
\end{figure}

Figure~\ref{F6} presents the computed frequency‐dependent response of a linear array of five superconducting flux qubits, both in the presence [(a,b)] and in the absence [(c,d)] of mutual inductive coupling. In panel (a), the amplitude of the response function $\lvert\chi(\omega)\rvert$ for the coupled array exhibits a pronounced main resonance at $\omega/\omega_{0}\approx0.18$, accompanied by smaller satellite peaks at higher normalized frequencies, which arise from collective excitation modes induced by mutual inductance energy $M I_s^2$. The corresponding phase response $\phi(\omega)/\pi$ in panel (b) shows multiple abrupt phase jumps: the primary $\pi$-step at the principal resonance is supplemented by additional discontinuities coinciding with each subsidiary resonance/antiresonance in the amplitude spectrum. In particular, even where the amplitude features are only weakly resolved, the phase response reveals unambiguously the nontrivial multi-state structure resulting from qubit-qubit coupling.

In contrast, panels (c) and (d) report the amplitude and phase response for five isolated qubits (with $-M I_s^2=10^{-6} \ \hbar\omega_0$). In this uncoupled limit, the amplitude spectrum in (c) reduces to a single Lorentzian resonance at $\omega/\omega_{0}\approx0.40$, and the phase in (d) displays the textbook $\pi$ phase roll-off characteristic of an isolated two‐level system. The direct comparison of (a,b) with (c,d) highlights the emergence of multiple resonances and phase discontinuities as unambiguous signatures of the qubit-qubit inductive coupling.

Because the phase response $\phi(\omega)$ is extremely sensitive to even weak secondary resonances or antiresonances, it constitutes a powerful experimental probe for the fine characterization of coupling‐induced mode structure in multi‐qubit circuits. The latter observation suggests that phase‐sensitive measurements could enable the precise determination of coupling strengths and collective eigenfrequencies, complementing conventional amplitude spectroscopy.\\

\begin{figure}
	\centering
	\includegraphics[width=1.0\textwidth]{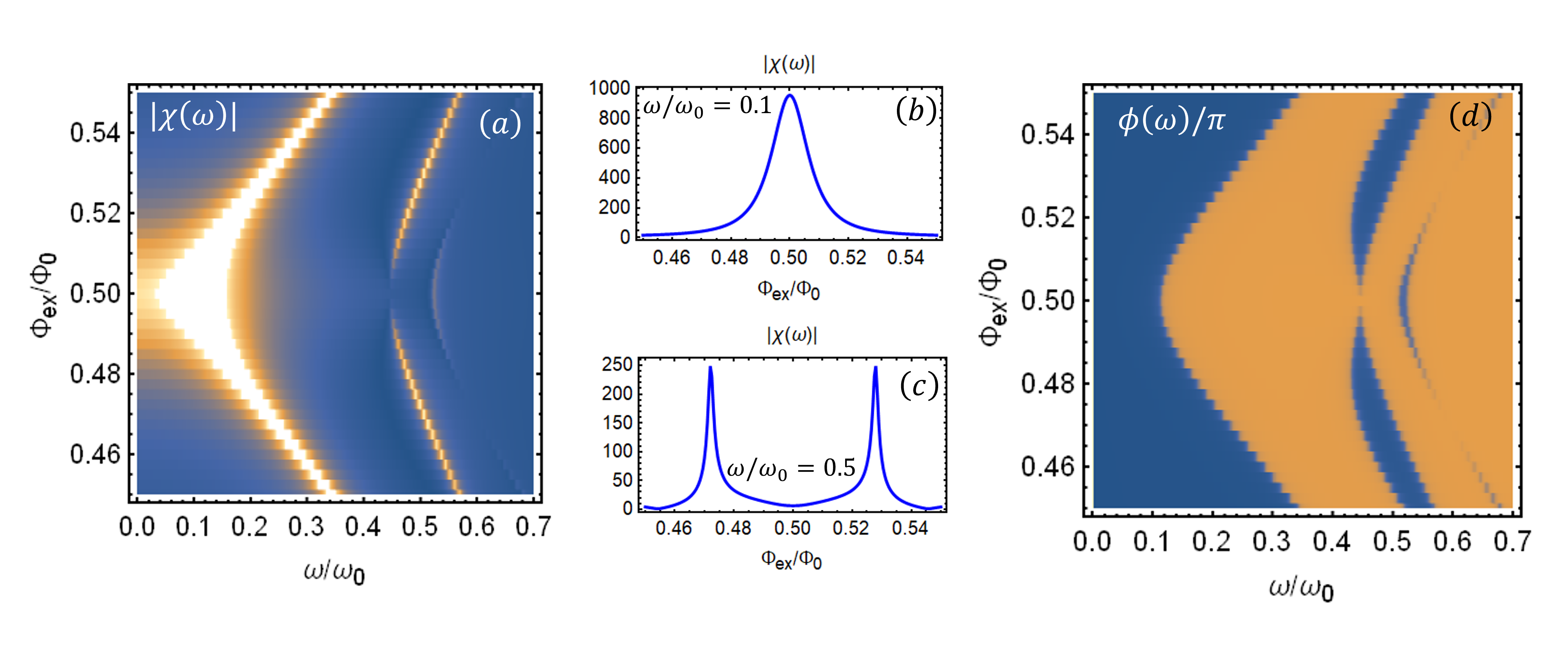}
	\caption{\textbf{(a)} Color–density map of the response function amplitude 
    $\lvert\chi(\omega)\rvert$, where $\chi(\omega)=\lvert\chi(\omega)\rvert\,e^{i\phi(\omega)}$, for a linear qubit array coupled to a transmission line, shown as a function of the normalized drive frequency $\omega/\omega_{0}$ and the applied flux $\Phi_{\mathrm{ex}}/\Phi_{0}$. Model parameters are set to: $\Delta=-MI_S^2=0.2 \ \hbar \omega_0$, $\eta=2.5 \cdot 10^{-3} \ \hbar \omega_0$, with $\omega_0=I_s \Phi_0/\hbar$. 
    \textbf{(b,c)} Vertical cuts of the response function amplitude, shown in panel (a), at fixed frequency ratios $\omega/\omega_{0}=0.1$ (b) and $\omega/\omega_{0}=0.5$ (c), illustrating the evolution from a single resonance peak to a double‑peak structure in the $\chi$ versus  $\Phi_{\mathrm{ex}}/\Phi_{0}$ curves. 
    \textbf{(d)} Color–density map of the response function phase $\phi(\omega)/\pi$ obtained by setting the model parameters as done in panel (a). In all density plots, darker shading corresponds to lower values of the plotted quantity.}
	\label{F7}
\end{figure}

Figure~\ref{F7} illustrates the flux‑ and frequency‑dependent response of the five‑qubit array in the non‑zero mutual‑inductance regime.  In panel (a) the color‑density map of the amplitude $\lvert\chi(\omega)\rvert$ is plotted as a function of the normalized drive frequency $\omega/\omega_{0}$ and the applied flux bias $\Phi_{\rm ex}/\Phi_{0}$.  One observes that the resonances position evolves when the external flux is detuned from the nominal working point of the qubit (i.e., $\Phi_{ex}/\Phi_0=0.5$), signaling the flux-tunability of the system.  

Panels (b) and (c) display vertical cuts through the amplitude map at fixed frequencies $\omega/\omega_{0}=0.1$ and $\omega/\omega_{0}=0.5$, respectively.  At low drive frequency [panel (b)], a single Lorentzian peak centered near $\Phi_{\rm ex}/\Phi_{0}=0.50$ gradually broadens and splits as the driving frequency increases. In particular, at higher frequency [panel (c)], the lineshape evolves into a clear double‑peak structure, reflecting the hyperbolic functional form of the excitation energy of the system.  

Finally, panel (d) shows the color‑density map of the response function phase $\phi(\omega)/\pi$ under the same parameter choice of panel (a).  The sharp phase discontinuities trace the same split branches seen in the amplitude map, and even in regions where amplitude contrast is weak, the phase exhibits distinct jumps at each resonance/antiresonance locus.  This pronounced phase sensitivity underscores the utility of phase-resolved spectroscopy for a precise mapping of the array dynamical response.  

\subsection{Dynamical Response of Linear and Cross-Shaped Arrays: A Comparison}
To ensure a fair comparison between linear and cross-shaped qubit arrays, we first recall that, in linear geometry, one can align the array parallel to the transmission line so that each qubit experiences an identical inductive coupling (the configuration used in previous analysis and reported in Fig. \ref{F2}, panel (c)). In contrast, the cross-shaped arrangement precludes a uniform coupling since only a peripheral qubit can be brought into substantial inductive coupling with the line. Thus, without modification of our previous coupling scheme, the two geometries would be subject to inherently different coupling strengths, confounding any direct comparison of their dynamical responses. To avoid this problem, we consider a modified coupling in which only the fifth qubit is directly affected by the influence of the transmission line (see Fig. \ref{F2}, panels (d) and (e)). The interaction Hamiltonian is then written as
\begin{equation}
H_I = \mathcal{M} I_S I(t) \sigma_z^{(5)} \equiv f(t) \sigma_z^{(5)},
\end{equation}
for both the linear and the cross-shaped array. This expression captures the dominant contribution to the interaction, as the mutual inductance coefficients between the transmission line and the qubits decay rapidly with distance. Therefore, the coupling to the fifth qubit, which is the closest to the line, provides a good approximation of the total interaction.

Consequently, the flux response of the qubit network to the external signal can be characterized by analyzing the susceptibility $\chi(\omega)$ in Eq.~(\ref{eq:chi}), with $A_S =\sum_i \sigma_z^{(i)}$ and $B_S =\sigma_z^{(5)}$.\\

\begin{figure}
	\centering
	\includegraphics[width=1.0\textwidth]{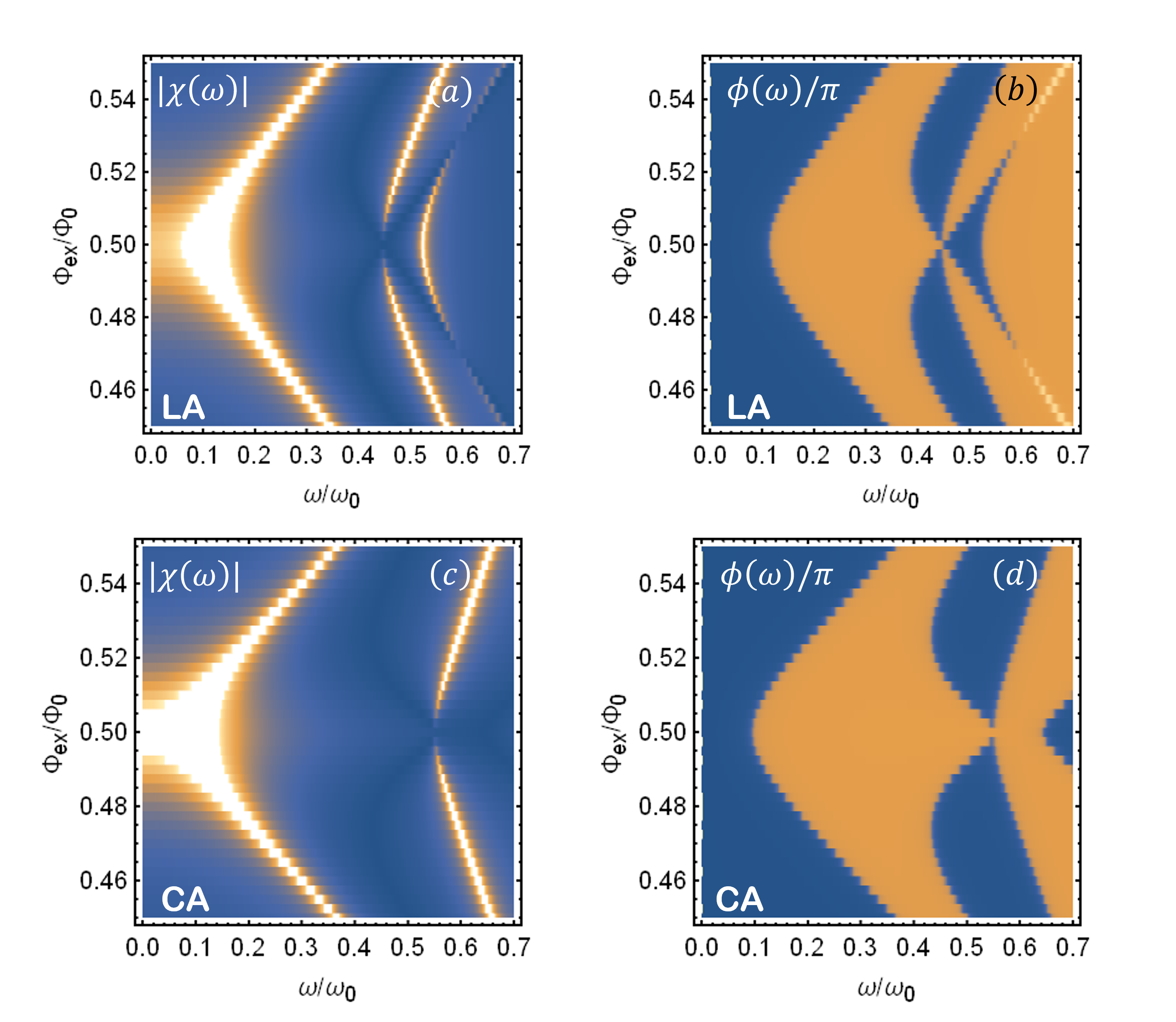}
	\caption{Amplitude (Panel (a)) and phase (Panel (b)) of the response function $\chi=|\chi|e^{i\phi}$ of a linear qubit array (LA) coupled to a transmission line via dominant inductive coupling with the 5th qubit. Panel (c) and (d) show the same quantities for a cross-shaped array (CA). The model parameters for all the panels are set to: $\Delta=-MI_S^2=0.2 \ \hbar \omega_0$, $\eta=2.5 \cdot 10^{-3} \ \hbar \omega_0$, with $\omega_0=I_s \Phi_0/\hbar$. In all panels, darker shading corresponds to lower values of the plotted quantity.}
	\label{F8}
\end{figure}

Figure~\ref{F8} displays a detailed comparison of the amplitude and phase of the response functions for linear (Panels (a)-(b)) and cross-shaped (Panels (c)-(d)) design of the qubit network. Specifically, Panels (a) and (b) illustrate the magnitude and phase, respectively, of the response function $\chi = |\chi| e^{i\phi}$ for the linear qubit array (LA) configuration, coupled inductively via the fifth qubit to a transmission line. Panels (c) and (d) present analogous quantities for the cross-shaped qubit array (CA), highlighting differences introduced by geometry. Notably, the amplitude response clearly exhibits resonant modes whose positions and widths are sensitive to the arrangement of the qubits, reflecting variations in coupling strength and mode hybridization. Similarly, the phase plots show abrupt shifts that signal resonance crossings or nodal structures stemming from the vanishing of matrix elements relevant to quantum transitions between the ground state and excited states.

\subsection{Dynamical Response of Disordered Arrays}

So far, we have considered arrays of identical flux qubits, characterized by uniform parameters. However, in realistic implementations, fabrication imperfections inevitably introduce random variations in qubit properties \cite{Fistul2022}. To assess the robustness of the system’s response under such imperfections, we investigate the effects of disorder in a linear array of flux qubits, which serves as a relevant study case to be compared to the response of the ideal system presented in Sec. \ref{sec: LA dynamic}. Disorder is incorporated in the model by replacing the loop critical currents $I_S^{(i)}$ and the tunneling energies $\Delta^{(i)}$ with $I_S(1 + \lambda_i)$ and $\Delta(1 + \mu_i)$, respectively, where $\lambda_i$ and $\mu_i$ are independent random variables uniformly distributed in the interval $[-0.1, 0.1]$. This choice reflects fabrication-induced parameter fluctuations up to $\pm 10 \ \%$ of the nominal values $I_s$ and $\Delta$. These modifications affect both the mutual inductive coupling between qubits and their interaction with the transmission line. In particular, the mutual coupling between qubits $i$ and $j$ is now weighted by the extra-factor $(1 + \lambda_i)(1 + \lambda_j)$, so that the Hamiltonian of the isolated array can be written as:
\begin{eqnarray}
H_{LA}^{(D)}=-\sum_{i=1}^{5} [(1+\lambda_i)\epsilon(f) \sigma_{z}^{(i)}+(1+\mu_i)\Delta \sigma_{x}^{(i)}]+\nonumber\\
+MI_{S}^{2} \sum_{i=1}^{4}(1+\lambda_{i})(1+\lambda_{i+1}) \sigma_{z}^{(i)} \sigma_{z}^{(i+1)},
\end{eqnarray}
while the coupling to the transmission line takes the modified form: 
\begin{equation}
H_I = \mathcal{M} I(t) \sum_i I_S^{(i)} \sigma_z^{(i)} \equiv f(t) \sum_i (1+\lambda_i)\sigma_z^{(i)}.
\end{equation}
Thus, the flux response of the disordered qubit network to the external signal can be characterized by analyzing the susceptibility $\chi(\omega)$ in Eq.~(\ref{eq:chi}), with $A_S = B_S = \sum_i (1+\lambda_i) \sigma_z^{(i)}$. Importantly, the response function $\chi$ is influenced by disorder not only through the explicit form of the operators $A_S$ and $B_S$, but also through disorder-induced modifications of the array's spectral properties, including both eigenstates and excitation energies.

\begin{figure}
	\centering
	\includegraphics[width=1.0\textwidth]{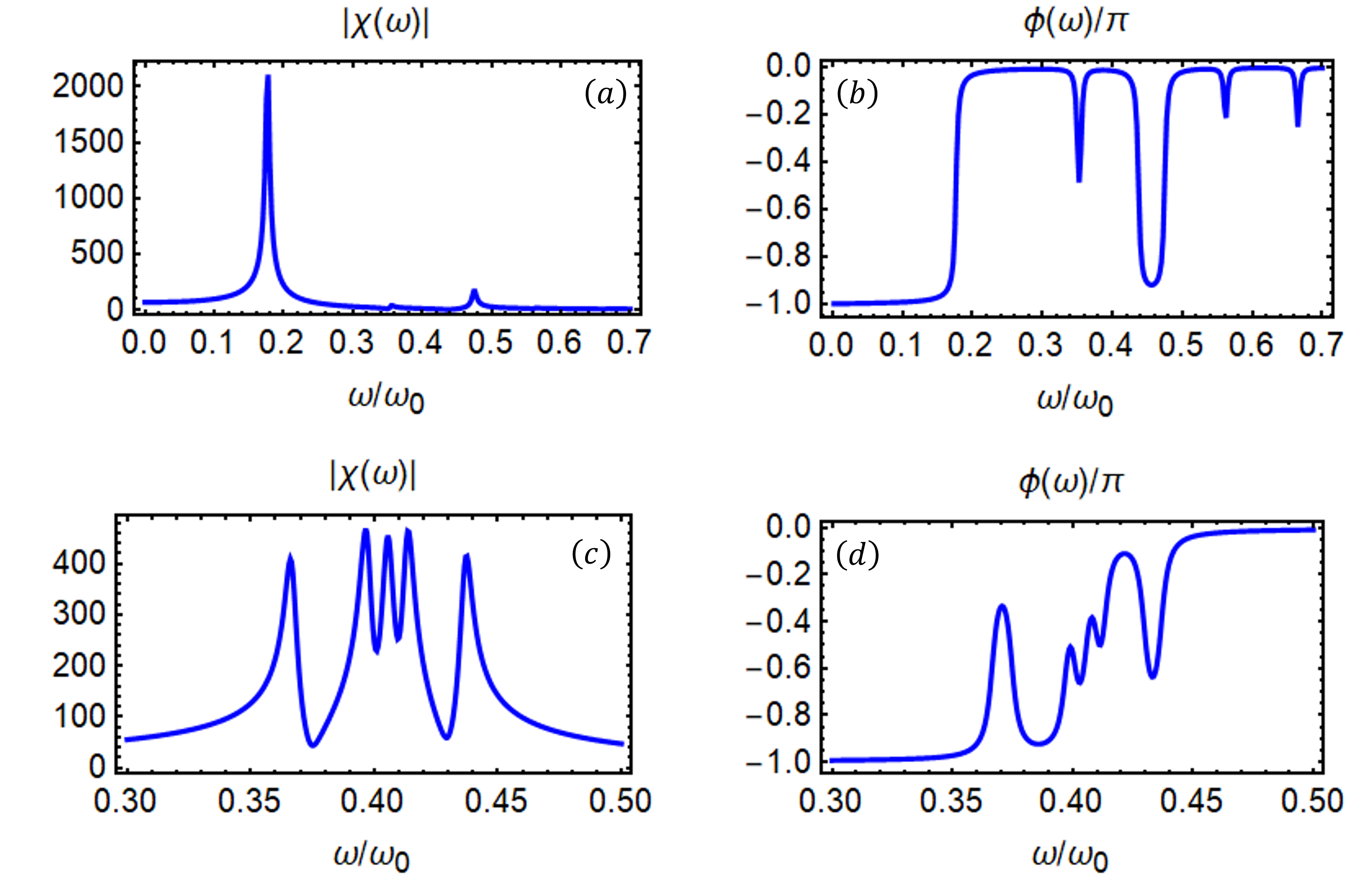}\\
        \includegraphics[width=1.0\textwidth]{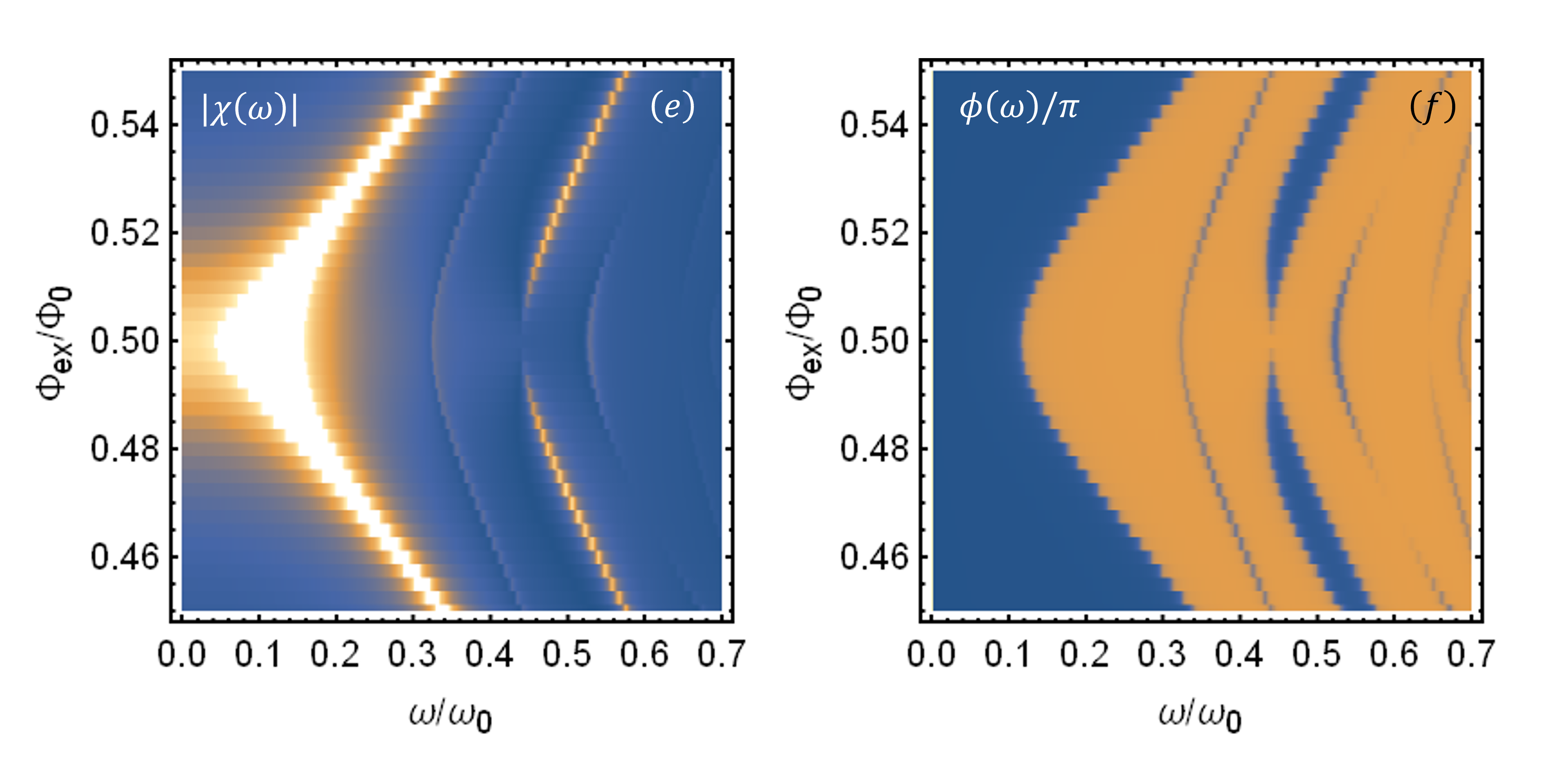}
	\caption{\textbf{(a)} $|\chi(\omega)|$ of a disordered linear array of five coupled qubits, plotted versus the normalized drive frequency $\omega/\omega_{0}$. The system parameters are: $\Phi_{ex}/\Phi_0=0.52$, $I_S^{(i)}=I_s(1+\lambda_i)$, $\Delta_i=\Delta(1+\mu_i)$, $\Delta=-MI_S^2=0.2 \ \hbar \omega_0$, $\eta=2.5 \cdot 10^{-3} \ \hbar \omega_0$, with $\omega_0=I_s \Phi_0/\hbar$. The random variables $\lambda_i$ and $\mu_i$ have been obtained as detailed in the main text. 
    \textbf{(b)} Corresponding response function phase $\phi(\omega)/\pi$. 
    \textbf{(c,d)} $|\chi(\omega)|$ and $\phi(\omega)/\pi$ for five isolated qubits, i.e. a system with mutual inductance coupling energy $-MI_s^2$ set to $10^{-6} \ \hbar \omega_0$, while taking the remaining parameters as in panels (a) and (b).
    \textbf{(e)} Color–density map of 
    $\lvert\chi(\omega)\rvert$ for a disordered linear qubit array coupled to a transmission line, shown as a function of the normalized drive frequency $\omega/\omega_{0}$ and the applied flux $\Phi_{\mathrm{ex}}/\Phi_{0}$. Model parameters are set as done in panel (a) and (b). \textbf{(f)} Color–density map of $\phi(\omega)/\pi$ under the same conditions as in (e). In all density plots, darker shading corresponds to lower values of the plotted quantity.}
	\label{F9}
\end{figure}
To illustrate the impact of disorder on the system's response, in Fig.~\ref{F9} we report the behavior of the susceptibility \(\chi(\omega)\), computed for a disordered linear array of five coupled qubits. In panels (a) and (b), we show the modulus and phase of the susceptibility as a function of the normalized drive frequency \(\omega/\omega_0\) for a fixed external flux \(\Phi_{ex}/\Phi_0 = 0.52\), and for a single realization of disorder. The presence of mutual inductive coupling between the qubits leads to a collective response, characterized by a dominant resonance peak and a rich phase profile. In contrast, panels (c) and (d) display the response function of the same disordered configuration under the same flux bias but with the mutual inductive coupling energy reduced by several orders of magnitude. In this weak-coupling regime, the collective nature of the response is suppressed, and the single broad resonance observed in panel (a) is replaced by a fragmented structure composed of five distinct peaks, corresponding to the individual transitions of the five qubits. Finally, panels (e) and (f) present the full frequency- and flux-resolved response for the same disorder realization as in panels (a) and (b). The color maps clearly reveal the evolution of the resonance features and phase response as a function of \(\Phi_{\text{ex}}\), highlighting the influence of the disorder parameters \(\lambda_i\) and \(\mu_i\) on the position and sharpness of the resonances. Notably, the qubit-qubit coupling tends to mitigate the impact of disorder on the system’s response function, promoting the emergence of collective features that closely resemble those found in the ideal, disorder-free linear array (see Fig.~\ref{F7}, panels (a) and (d)).

\section{From Detection to Information Processing: Flux-Qubit Arrays as Quantum Reservoirs}
\label{sec:8}
In addition to enabling quantum-limited sensing, superconducting qubit arrays can be harnessed as computational systems that actively support the sensing task itself. In this dual role, a flux-qubit array not only responds to external electromagnetic stimuli with high sensitivity, but can also process and transform these signals through its intrinsic nonlinear dynamics. This perspective opens the possibility of integrating in-sensor computation with quantum-limited detection, thus leveraging the same physical platform for both measurement and information processing.

To implement such hybrid functionality, a promising approach is offered by the paradigm of Quantum Reservoir Computing (QRC), in which the quantum dynamics of the sensing device is used simultaneously as a quantum reservoir, capable of extracting temporal features and supporting learning-based inferences.

The QRC paradigm \cite{Fujii2017,Abbas2024,Zhu2024,
Mujal2022,Sannia2024,Mifune2024,Xia2023,Martínez2023,Gotting2023,LoMonaco2024,Vetrano2024,Palacios2024,Monzani2024,Ivaki2024,Ahmed2024,Kobayashi2024a,
Kornjaca2024} merges the potential of quantum systems \cite{Shor1994, Grover1996, Nielsen2010, Montanaro2016, Preskill2018, Mastroianni2022, Mastroianni2023, Mastroianni2024, Consiglio2024, DeLorenzis2024, Boneberg2023,Damore2025} with the classical reservoir computing approach\cite{Maass2002, Jaeger2004}, which efficiently processes time-dependent signals by embedding them in the rich and high-dimensional dynamics of a fixed reservoir.

A necessary condition to effectively implement this program is that the driven dynamics of the qubit array, as studied in the previous sections, exhibits a sufficient degree of richness and non-linearity to perform non-trivial temporal processing tasks. 

To assess this capability, we study flux-qubit arrays, when subject to an external frequency-modulated signal, and show that they can be used to perform prediction of a chaotic time series, which is a standard benchmark task in the QRC framework. We will also demonstrate that the topology and parameters of the qubit network can be optimized to enhance the system performance.

In the following, we detail the implementation of our scheme. Let \(\{s_k\}\) be a generic input sequence normalized to the interval \([0,1]\). We encode the \(k\)-th input value by modulating the drive frequency of the external signal according to 
\[
f_k(t) = \mathcal A \,\sin\bigl(\omega_k\,t\bigr), 
\qquad 
\omega_k = \omega_{\min} + s_k\,(\omega_{\max} - \omega_{\min}), 
\]
where \(\mathcal A\) is a fixed amplitude and \(\omega_k\) varies linearly between \(\omega_{\min}\) and \(\omega_{\max}\) as \(s_k\) runs from 0 to 1.
For each input \(s_k\), the flux‐qubit network is initialized in the ground state of the unperturbed Hamiltonian \(H_0\), denoted by $\ket{0}$. 
 The chosen set of observables \(\{ \hat{O}_\alpha \}\) (as in Eq.~(23) ) are accumulated in a time‐multiplexed scheme at \(  n_t \) equally spaced instants 
\[
t_j = j\,\frac{ t_{max} }{n_t-1}, 
\quad j=0,1,\dots,n_t-1,
\]
yielding an expectation‐value vector \(\vec{m}_k\in\mathbb{R}^{d \cdot n_t+1}\), where \(d\) is the number of observables per time step. To incorporate the memory of previous inputs, we follow the scheme presented in \cite{Settino2024}. We define the reservoir state \(\vec{r}_k \in \mathbb{R}^{l_r}\) through
\[
\vec{r}_k = \gamma\,\hat{S}^{\,n_S}\,\vec{r}_{k-1} + \hat{B}^{\,l_r}\,\vec{m}_k,
\]
where \(\gamma\in(0,1)\) controls the memory decay, \(\hat{S}^{\,n_S}\) is the cyclic shift operator of dimension \(l_r\) that permutes each component forward by \(n_S\) positions,
\( [\hat{S}^{\,n_S}]_{ij} = \delta_{(i+n_S)\bmod l_r,\,j}, \)
and \(\hat{B}^{\,l_r}\) is a lengthening operator that embeds \(\vec{m}_k\) into an \(l_r\)-dimensional vector by interleaving zeros, ensuring that \(\vec{r}_k\) has fixed dimension \(l_r\). The reservoir state is initialized as \(\vec{r}_0 = \mathbf{0}\).

After collecting reservoir states \(\{\vec{r}_k\}_{k=1}^{N_{\mathrm{tr}}}\) over \(N_{\mathrm{tr}}\) training inputs, we assemble the following matrix:
\[
\mathbf{R} 
= 
\begin{bmatrix}
\vec{r}_1^\mathrm{T} \\ 
\vec{r}_2^\mathrm{T} \\ 
\vdots \\ 
\vec{r}_{N_{\mathrm{tr}}}^\mathrm{T}
\end{bmatrix}
\in\mathbb{R}^{N_{\mathrm{tr}}\times l_r},
\]
and define the target vector \(\vec{y} = [\,y_1,\,y_2,\,\dots,\,y_{N_{\mathrm{tr}}}\,]^\mathrm{T}\), where \(y_k\) is chosen according to the task. The linear readout weights \(\vec{w}\in \mathbb{R}^{l_r}\), which contain the relevant information about the problem, are then obtained by ordinary least squares procedure in the form:
\[
\vec{w} = \bigl(\mathbf{R}^\mathrm{T}\mathbf{R}\bigr)^{-1} \,\mathbf{R}^\mathrm{T}\,\vec{y}.
\]
Once the weight vector \(\vec{w}\) is found, the predicted output for step \(k\) is simply given by \(\tilde{y}_k = \vec{w}^\mathrm{T}\,\vec{r}_k\).\\

The general QRC scheme presented above is now used to perform a benchmark prediction task. For this purpose, we generate a chaotic time series derived from the Mackey-Glass (MG) delay differential equation:

\begin{equation}
\frac{ds(t)}{dt} = \beta \frac{s(t - \tau)}{1 + s(t - \tau)^n} - \Gamma s(t).
\end{equation}

The differential equation is numerically solved and the solution is sampled with a time step \( \delta t \), to define a discrete sequence. 

The simulation parameters used here are fixed as done in Ref.~\cite{Settino2024,Sannia2024}, i.e.   
$\beta=0.2$, $ \Gamma=0.1$, delay time $\tau=17$ (chaotic behavior emerges for $\tau \ge 17$), non-linearity parameter $n=10$, and the discretization time $\delta t=3.0$. In the training phase, we implement a one-step forward prediction task, with $y_k=s_{k+1}$. In the test phase, starting from the last training time, a longer-time prediction is obtained by multiple one-step predictions.  

To evaluate the performance of our reservoir model, we define a Valid Prediction Time (VPT), according to Ref. \cite{Wudarski2023,Settino2024}. 
The VPT is defined as the maximum time \( T \) over which the predicted trajectory \( \hat{y}(t) \) remains within a prescribed error threshold from the true trajectory \( y(t) = s(t) \). This condition can be mathematically formulated as follows:

\begin{equation}
VPT = \max \left\{ T : \forall t \leq T, \left(\frac{\tilde{y}(t)-y(t)}{\sigma}\right)^2 < \varepsilon^2, \right\}
\end{equation}
where \( \epsilon=0.3 \) is an error tolerance and $\sigma$ is the standard deviation of the time series. We express VPT in integer units of the discretized time.   

\begin{figure}
    \centering
    \includegraphics[width=1.\linewidth]{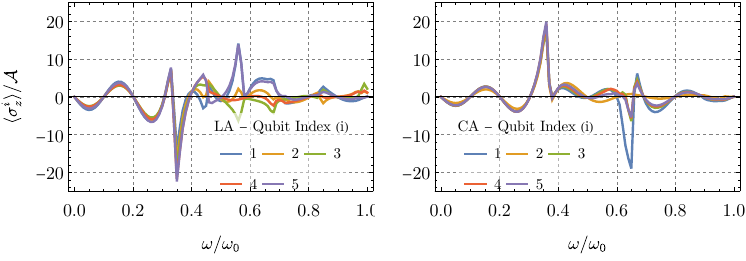}
    \caption{
        Expectation values \(\langle \sigma_z^{(i)} \rangle\) for each of the five qubits as a function of the drive frequency \(\omega/\omega_0\), evaluated at a fixed time \(t =  2\pi/\omega_{\min}\) for LA configuration (left panel) and CA configuration (right panel). Parameters: normalized flux \(f = \Phi_{\mathrm{ex}}/\Phi_0 = 0.45\); inhomogeneous tunneling energies \(\Delta_i\) linearly distributed in \(\Delta [1-\delta,1+\delta]\), with \(\delta = 0.1\); inductive coupling energy \(- M I_S^2 = 0.2\,\hbar\omega_0\), with
        \(\hbar \omega_0 = I_S\Phi_0\).
    }
    \label{fig:response}
\end{figure}

Before analyzing the prediction performances, we show the non-linear response of the qubit loop currents to a frequency-modulated drive.  
To enhance the richness of the system's response, we set the tunneling energies \(\Delta_i \) as linearly spaced quantities belonging to the interval \( \Delta [1-\delta, 1+\delta] \), being the parameter $\delta$ related with the inhomogeneity of the tunneling energy.
Figure~\ref{fig:response} shows the expectation values \(\langle \sigma_z^{(i)} \rangle\) for each of the five qubits as a function of the driving frequency \(\omega/\omega_0\), evaluated at a fixed measurement time. The left panel corresponds to the linear array (LA) with homogeneous coupling to the transmission line (see Fig. \ref{F2} (c) ), while the right panel reports the results for the cross-shaped array (CA) (see Fig. \ref{F2} (e)).
As evident from the plots, the system exhibits a highly non-linear dependence of the loop currents on the drive frequency, with multiple resonant peaks and dips. Notably, the small inhomogeneity introduced in the tunneling energies $\Delta_i$ leads to different responses across the individual qubits, which is particularly visible near resonance. This heterogeneity enhances the learning performance of the reservoir, providing a variety of features that improve the expressivity of the system. Moreover, in the cross-shaped geometry, the enhanced symmetry reduces the variety of responses.  These features form the basis for the frequency-encoded QRC strategy employed in subsequent prediction experiments.

\begin{figure}
    \centering
    \includegraphics[width=0.9\linewidth]{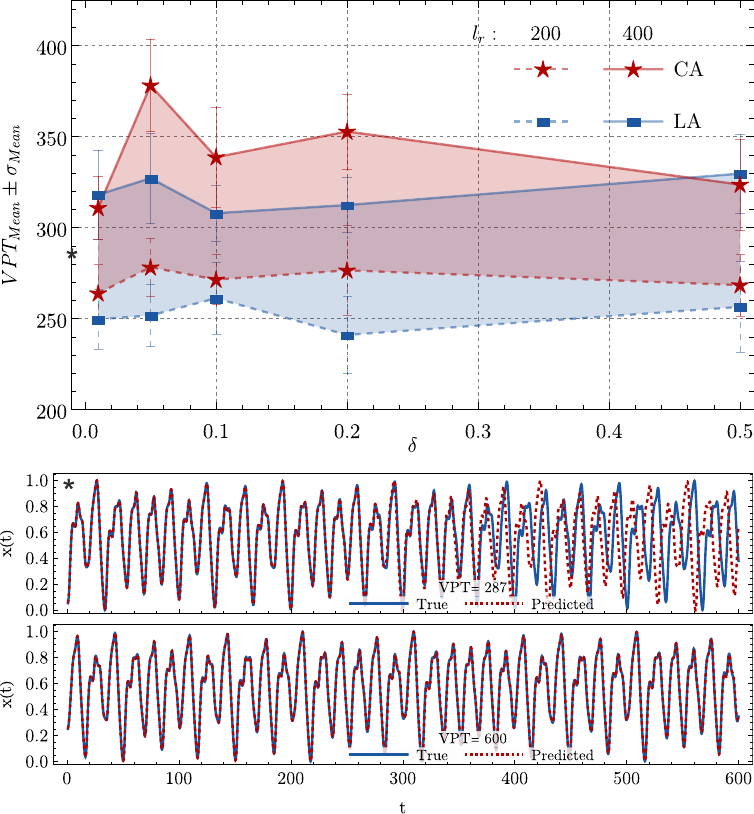}
\caption{
        (Top panel) Valid Prediction Time (VPT) as a function of the inhomogeneity parameter \(\delta\) for the Mackey–Glass prediction task. Results are shown for reservoir sizes \(l_r = 200\) and \(l_r = 400\), and for LA and CA configurations. QRC parameters: \(\gamma = 0.6\), \(n_S = 1\), \(n_t = 6\), \(t_{\max} =  2\pi/\omega_{\min}\), \(\omega_{\min} = 0.2\,\omega_0\), \(\omega_{\max} = 0.6\,\omega_0\), \(\mathcal{A} = 0.001\,\hbar\omega_0\).  (Bottom panels) Prediction examples for the CA configuration (\(\delta = 0.1\), \(l_r = 400\)): with \(\mathrm{VPT}=287\) (poor generalization), and with \(\mathrm{VPT}>600\) (excellent agreement). The true series is shown as solid blue line, while the prediction is shown as dashed red line.
    }
    \label{fig:QRC}
\end{figure}

Fig.~\ref{fig:QRC} focuses on the one-step-ahead MG prediction task. In the top panel of Fig.~\ref{fig:QRC}, we show the VPT as a function of the inhomogeneity parameter $\delta$ for two different reservoir dimensions $l_r$, for both the cross-shaped array and the linear array configuration. 
Firstly, we observe that a small inhomogeneity improves the network's performance; this improvement is more pronounced in the cross-shaped configuration. For all the $\delta$ values, the CA configuration gives an advantage over the LA configuration. This behavior depends on the inhomogeneous coupling with the transmission line, which is peculiar to the CA configuration. As expected, the increase of the reservoir dimension improves the VPT, because the prediction of the MG series is enhanced by storing more past inputs in the network. 

In the bottom panels of Fig.~\ref{fig:QRC}, we present two representative single-shot realizations of the predicted time series generated by the cross-shaped array (CA) reservoir.

The first panel shows a suboptimal prediction instance, characterized by a VPT of \(287\). In this case, the predicted trajectory initially follows the true Mackey-Glass signal but gradually diverges. In contrast, the second panel displays a realization in which the reservoir maintains excellent agreement with the target signal over the entire duration of the test window. Here, the VPT exceeds the maximum evaluation time, indicating that the reservoir dynamics has effectively captured the structure of the time series and generalized well beyond the training window. 

These examples demonstrate not only the potential of the CA configuration for robust time-series prediction but also revile the sensitivity of performance to specific levels of inhomogeneity and reservoir parameter choices. They underscore the importance of both geometric design and controlled inhomogeneity in enhancing the computational capabilities of the proposed quantum reservoir systems.

These findings support the feasibility of a dual-function architecture, in which the flux-qubit array serves both as a quantum-limited electromagnetic sensor and as a quantum reservoir capable, for instance, of solving inverse problems through in-sensor information processing. By monitoring accessible observables of the array, one may reconstruct the input signal, effectively integrating quantum sensing and quantum reservoir computing. This perspective defines a promising strategy for integrating signal detection and processing in a unified quantum hardware platform.

\section{Conclusions}
\label{sec:9}
In conclusion, we have systematically analyzed the role of network topology in determining the spectral and flux-response properties of superconducting qubit networks. Through exact diagonalization and linear response theory, we demonstrated that the geometric arrangement of inductively coupled qubits substantially influences both static and dynamic system properties. Specifically, we found that the cross-shaped topology significantly enhances circulating currents and the generated magnetic flux compared to linear configurations. These results underline the potential of topological engineering in optimizing superconducting qubit networks, highlighting its practical relevance for quantum computing and high-sensitivity quantum sensing applications.\\
Our findings provide a concrete framework for future experimental and theoretical studies aimed at exploiting topology to enhance quantum device performance. In particular, the possibility of leveraging the same flux-qubit network both as a quantum-limited sensor and as a quantum reservoir defines a new paradigm for signal reconstruction and adaptive sensing, where the observed response variables provide the basis for the solution of the inverse problem via reservoir computing techniques. Such integrated quantum platforms could play a pivotal role in future quantum-enabled technologies for sensing, learning, and control.\\
Furthermore, the theoretical description presented in this paper is of considerable importance for the quantum integration of superconducting quantum networks. Recent experimental advances achieved by some of us, in which collective quantum effects were observed and characterized in superconducting architectures of diverse geometries \cite{Delia2022,Gatti2023}, reinforce this perspective. These parallel efforts establish an encouraging ground for validating and extending the present theory and for guiding the design of novel quantum devices based on engineered network topology.\\ 
In particular, the possibility of using the many-body manifold spanned by the collective ground state \(\lvert G\rangle\) and the first excited state \(\lvert E\rangle\) of a cross-shaped array to encode a logical qubit represents a promising perspective. Such an encoding can passively suppress uncorrelated decoherence and thereby prolong coherence times.\\ 
In the future, experiments will be essential to chart the ranges of temperature, fabrication tolerance, and noise correlation in which this protection delivers a tangible advantage over single-qubit platforms complemented by active error correction.

\section{Acknowledgment}
The research activity of BR, PS and ML received partial support from the PNRR  MUR projects National Quantum Science and Technology Institute - NQSTI (Partenariato Esteso 04: Scienze e Tecnologie Quantistiche PE0000023). ML received partial support by National  Center  for  HPC,  Big  Data  and Quantum  Computing  –  HPC  (Centro Nazionale 01-CN0000013). BR and PS are also supported by Associazione EUDORA APS ATS under the Project Quantum Nexus.\\
JS acknowledges the contribution from PRIN (Progetti di Rilevante Interesse Nazionale) TURBIMECS, grant n. 2022S3RSCT CUP H53D23001630006, and PNRR MUR project PE0000023 - NQSTI through the secondary projects “ThAnQ” J13C22000680006.\\ 
The research of GGL is supported by the postdoctoral fellowship program of the University of Lleida.\\
The research activity of FR received support from the PNRR MUR project PE0000023 – NQSTI, through the cascade-funded projects SPUNTO (CUP E63C22002180006, D.D. MUR No. 1564/2022) and TOPQIN (CUP J13C22000680006, D.D. MUR No. 1243/2022).\\
The authors gratefully acknowledge Matteo Cirillo, whose pioneering experimental work inspired their investigation of topological effects in superconducting networks. They also acknowledge Carmela Bonavolontà and Alfonso Maiellaro for insightful discussions on the subject of this study and Davide Buono for his valuable technical support.

% \cite{nocite_entry}
% \nocite{*}
\bibliographystyle{ieeetr}    
% scegli uno stile: plain, unsrt, alpha, abbrv…
\bibliography{bibQuBit}
\end{document}